\documentclass[twocolumn,showpacs,preprintnumbers,amsmath,amssymb]{revtex4}

\usepackage{graphicx}
\usepackage{dcolumn}
\usepackage{bm}

\newcommand{\ket}[1]{|\,#1\,\rangle}
\newcommand{\bra}[1]{\langle\,#1\,|}
\newcommand{\braket}[2]{\langle {#1}|{#2}\rangle}

\newcommand{\BesselJ}[1]{{\rm J}_{#1}}

\newcommand{\NAME}[1]{{#1}, }
\newcommand{\REVIEW}[4]{#1 {\bf #2}, {#3} (#4)}
\newcommand{\BOOK}[4]{{\it #1} (#2, #3, #4)}
\newcommand{\kbar}{k\hspace{-1.1ex}\raise0.3em\hbox{-}}
\begin{document}

\preprint{APS/123-QED}

\title{State reconstruction of the kicked rotor}

\author{M. Bienert, F. Haug and W. P. Schleich\\}
 \affiliation{Abteilung f\"ur Quantenphysik, Universit\"at Ulm, Albert--Einstein--Allee 11, D--89069 Ulm, Germany}
\author{M. G. Raizen}
\affiliation{Center for Nonlinear Dynamics and Department of Physics\\The University of Texas at Austin, Austin, TX 78712-1081}
\date{\today}

\begin{abstract}
We propose two experimentally feasible methods based on atom interferometry to measure the quantum state of the kicked rotor.
\end{abstract}


\maketitle
Atom optics \cite{bib:review} has become a very active experimental testing ground for quantum chaos \cite{bib:chaosbooks}. Such experiments \cite{bib:atomopticsandchaos} have investigated dynamical localization, quantum resonances and quantum dynamics in a regime of classical anomalous diffusion. So far, these studies have concentrated on the probability distribution rather than the full quantum state determined by amplitude {\it and} phase. \\
State reconstruction of simple quantum systems \cite{bib:staterec} has become a standard routine in labs. Such systems range from the motional state of a single ion in a trap \cite{bib:ionintrap} via tomography \cite{bib:tomography} of a single photon state \cite{bib:schiller} or atomic beams \cite{bib:mlyneknature} to quantum state holography \cite{bib:holography} of a Rydberg electron \cite{bib:rydbergatom}. Moreover, many theoretical suggestions \cite{bib:star, bib:selfinterference} exist. In the present paper we propose two methods to reconstruct the quantum state of the kicked rotor. 

Our proposal is an application of atom interferometry to quantum chaology \cite{bib:darcy}: We consider an atom whose motional degree of freedom is determined by a classically chaotic Hamiltonian. Entanglement between the internal and external dynamics allows us to measure the quantum state of the motion. The kicked rotor \cite{bib:kickedrotor} in its realization of a kicked particle \cite{bib:moore} serves as an illustration of our scheme. We emphasize that the present state--of--the--art of experimental techniques suffices to perform this experiment.

\begin{figure}
\includegraphics[width=8.5cm]{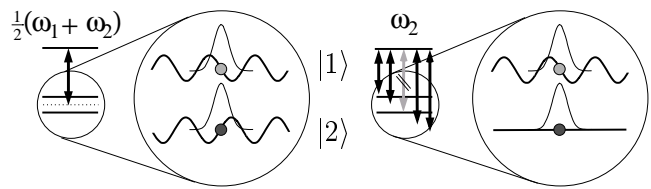}
\caption{
Methods of self--interference {\it (left)} and holography {\it (right)} to reconstruct the wave function of the kicked rotor. The method of self--interference considers a three--level atom with a laser field detuned half between the two hyperfine levels $\ket 1$ and $\ket 2$. The periodic potentials due to the light shifts experienced by the individual levels are out of phase. The method of holography uses a phase modulator to put sidebands on the transition frequency whereas the frequency $\omega_2$ itself is filtered out.
This symmetric detuning around $\omega_2$ leads to a constant potential for the lower state $\ket{2}$ but a periodic potential for the upper state $\ket{1}$. In both methods atoms in the state $\ket 1$ play the role of the kicked rotor whereas the atoms in state $\ket 2$ serve as our reference. 
}
\label{fig:model}
\end{figure}

Both reconstruction methods rely on controlling the dynamics of the atomic states $\ket 1$ and $\ket 2$ associated with two energy levels by the use of a laser field that couples the two levels with an excited electronic state (Fig.~\ref{fig:model}). This coupling influences the translational motion of the atom in the standing wave of the laser. Initially, a laser pulse prepares the atoms in a coherent superposition of state $\ket 1$ and $\ket 2$ establishing in this way a reference phase. It is followed by a sequence of $N$ pulses kicking the system. In the method of {\it self--interference} we use a classical electromagnetic standing wave that is detuned to the middle of the atomic transition. In this case the periodic potential felt by the atoms in state $\ket{1}$ is shifted by half a wavelength with respect to the potential corresponding to state $\ket{2}$. In the {\it holographic method} the standing wave is prepared in such a way as to only influence the atom in the upper state $\ket{1}$. Consequently, an atom in the lower state $\ket{2}$ propagates freely. The readout is common to both methods. A laser pulse shifts the momentum wave function of the lower state in order to measure the phase at the individual momenta.

We consider the quantum mechanical motion of an atom of mass $m$, characterized by coordinate $x$ and momentum $p$.
This motion is driven by appropriately tailored $\delta$-function kicks. They serve two purposes: On one hand they create the kicked rotor, on the other hand, they provide the readout of the wave function. 
The state $\ket{\phi(T_+)}$ of the center--of--mass motion immediately after a $\delta$-function kick described by the Hamiltonian
\begin{equation}
\hat H_{\delta} = \frac{\hat p^2}{2m} + V(\hat
x)\delta(t-T) 
\label{eq:displacementhamiltonian}
\end{equation}  
is related to the state $\ket{\phi(T_-)}$ just before the kick by a phase determined by the potential $V(x)$, that is
\begin{equation}
\ket{\phi(T_+)} = \exp\left[-\frac{i}{\hbar}V(\hat x)\right]\ket{\phi(T_-)}.
\label{eq:kickeq}
\end{equation} 
In an experimental realization the potential results from the interaction of the atomic dipole with the electromagnetic field in a given mode.
A standing wave of wavenumber $k_0/2$ creates a periodic potential $V(x)=\kappa \sin(k_0 x)$ leading \cite{bib:atomopticsandchaos} to the Hamiltonian 
\begin{equation}
\hat H = \frac{\hat p^2}{2m} + \kappa\sin{(k_0\hat
x)}\delta_T(t)
\label{eq:kickedrotorhamiltonian}
\end{equation}
of the kicked rotor \cite{bib:kickedrotor}. 
Here we assume a sequence of pulses $\delta_T(t)\equiv\sum_{n=-\infty}^\infty\delta(t-nT)$ of period $T$ and $\kappa$ denotes the kicking strength of the light field.

The Schr\"odinger equation together with the Hamiltonian, Eq.~(\ref{eq:kickedrotorhamiltonian}), determines the state $\ket{\phi(t)}$ at time $t$. Due to the sequence $\delta_T(t)$ of pulses it is convenient to consider the recurrence relation 
\begin{equation}
\ket{\phi_N}=\exp\left[-i\frac{\kappa}{\hbar}\sin(k_0 \hat x)\right]\exp\left[-i\frac{\hat p^2}{2 m}\frac{T}{\hbar}\right]\ket{\phi_{N-1}}
\end{equation}
connecting the state $\ket{\phi_N}\equiv|\phi(N T)\rangle$ immediately after the $N$--th kick with the state $\ket{\phi_{N-1}}$ after the kick $N-1$.
Throughout the paper we focus on the momentum wave function $\phi(p, t)\equiv\langle p|\phi(t)\rangle$. With the help of the expansion $\exp\left(i z \sin\theta\right)=\sum_l\BesselJ{l}(z)\exp\left(i l \theta\right)$ in terms of Bessel functions $\BesselJ{l}$ the momentum probability amplitude $\phi_N(p)\equiv\langle p|\phi_N\rangle$ obeys the mapping
\begin{equation}
\phi_N(p) = \sum_{l=-\infty}^{\infty}\BesselJ l\left(-\frac{\kappa}{\hbar}\right)e^{-i\beta(p-lp_0)}\phi_{N-1}(p-l p_0).
\end{equation}
Here we have introduced the phase $\beta(p)\equiv p^2 T/(2 m \hbar)$ quadratic in the momentum and the abbreviation $p_0\equiv\hbar k_0$. 
When we iterate this recurrence relation $N$ times we obtain a wave function with a rather complicated behavior of phases. This feature is due to the quadratic phase factor $\beta(p)$ arising from the free time evolution between the kicks. 

In order to measure the phase of a wave function we need an interferometric scheme with a reference phase. In our proposal we use the superposition $(\ket 1 + \ket 2)/\sqrt{2}$ between the internal states $\ket 1$ and $\ket 2$ of the atom. The level $\ket{1}$ is associated with the motion of the kicked rotor whereas $\ket{2}$ provides a reference. The initial state 
\begin{equation}
\ket{{\Psi}(t=0)} = \frac{1}{\sqrt2}\left[\ket 1 + \ket 2\right] \ket{\phi_0} 
\end{equation}
of the complete system consists of the internal states and the state $\ket{\phi_0}$ of the center--of--mass motion.

In our reconstruction method the two internal states undergo different dynamics governed by the Schr\"odinger equation
\begin{equation}
i\hbar \frac{\partial}{\partial t}\ket{\Psi(t)}= \left[\hat H\ket 1\bra 1 +\hat {\mathcal H}\ket 2\bra 2\right]\ket{\Psi(t)}
\end{equation}
leading to the state
\begin{equation}
\ket{{\Psi}(t)} = \frac{1}{\sqrt{2}}\left[\ket 1 \ket{\phi(t)}+\ket
2\ket{\varphi(t)}\right].
\label{eq:state1}
\end{equation}
Indeed, atoms in the upper state feel the dynamics of the kicked rotor and are described by the state 
$
\ket{\phi(t)}
$ obtained by propagating $\ket{\phi_0}$ with the Hamiltonian $\hat H$. In contrast, atoms in the reference state $\ket 2$ experience the Hamiltonian $\hat {\mathcal H}$ giving rise to the state 
$
\ket{\varphi(t)}.
$

\begin{figure}
\includegraphics[width=8.5cm]{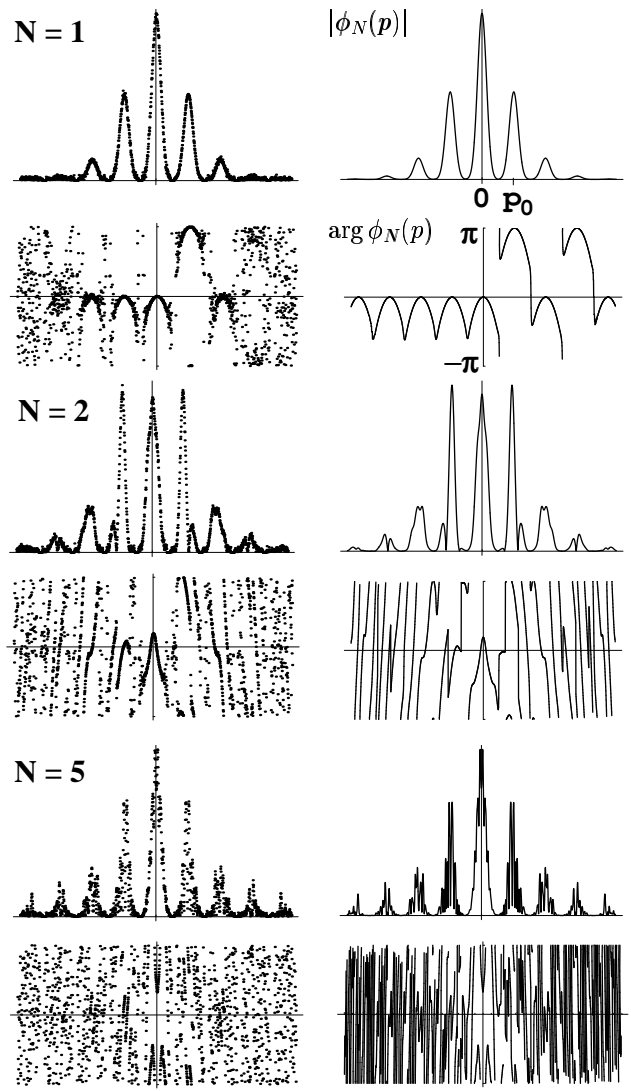}
\caption{
Monte--Carlo--Simulation {\it (left column)} of the holographic method for the state reconstruction of the kicked rotor after $N=1, 2$ and $5$ kicks. For each value of $N$ we represent the state by the amplitude {\it(top)} and the phase {\it(bottom)} of the momentum wave function. For comparison we also show the exact functions {\it (right column)}. For the reconstruction of the momentum wave function we use a grid of 4096 points and $M=10^6$ simulated measurement events.
}
\label{fig:recsim}
\end{figure}

We now turn to the readout of the wave function. For this purpose we apply after $N$ kicks a final $\delta$--function kick with a linear potential $V_P(x) = P x$ to the atom in $\ket 2$, that means a final laser pulse shifts the momentum wave function of state $\ket 2$ in order to measure the wave function of state $\ket 1$.  According to Eq. (\ref{eq:kickeq}) the state immediately after this kick reads 
\begin{equation}
\ket{\Psi_N}  \equiv \frac{1}{\sqrt{2}}\left[\ket 1 \ket{\phi_N} +  \ket 2 e^{-i P\hat x/\hbar}\ket{\varphi_N} \right]
\label{eq:state2}
\end{equation}
where $\ket{\varphi_N}\equiv\ket{\varphi(NT)}$.

Hence, the probability 
$
W_N^{(\theta)}(p; P) =
\left|\bra{j(\theta)}\bra{p}{\Psi_N}\rangle\right|^2
$
to find the atom in the superposition $\ket{j(\theta)} = \frac{1}{\sqrt{2}}\left[\ket 1+e^{-i\theta}\ket 2\right]$ with momentum $p$ takes the form
\begin{eqnarray}
W_N^{(\theta)}(p; P) &=& \frac{1}{2}\Big\{W_N(p)+{\mathcal W}_N(p+P)\nonumber\\
&&+{\rm
Re}\left[\phi_N(p)\varphi_N^\ast(p+P)e^{-i\theta}\right]\Big\}.
\label{eq:theWdist}
\end{eqnarray}
Here we have made use of the relation $\langle p|e^{-i P\hat x/\hbar} |\varphi_N\rangle=\langle
p+P|\varphi_N\rangle\equiv\varphi_N(p+P)$. Moreover, the distributions $W_N(p)=
\left|{\phi_N}(p) \right|^2/2$ and ${\mathcal W}_N(p)=\left| \varphi_N(p) \right|^2/2$ are the probabilities to find the atom in the upper state or in the reference state with the momentum $p$, respectively.

In order to reconstruct the wave function $\phi_N(p)$ we need to measure the probability distribution $W_N^{(\theta)}(p; P)$ for two angles $\theta=0$ and $\theta= \pi/2$ together with the momentum distributions $W_N(p)$ and ${\mathcal W}_N(p)$. 
With the help of Eq.~(\ref{eq:theWdist}) for these angles $\theta$ we can express the product
\begin{equation}
\phi_N(p)\varphi_N^\ast(p+P)={\mathcal M}_N(p;P)
\label{eq:theproduct}
\end{equation}
in terms of the sum
\begin{eqnarray}
{\mathcal M}_N(p;P)&=&2W_N^{(0)}(p; P)+2 i W_N^{(\frac{\pi}{2})}(p;P)\nonumber\\
&&-(1+i)\left[W_N(p)+{\mathcal W}_N(p+P)\right]
\label{eq:thesum}
\end{eqnarray}
of measured distributions.
The inversion formula, Eq. (\ref{eq:theproduct}), is the central tool for the reconstruction of the kicked rotor's wave function.

We now illustrate our reconstruction scheme by discussing two special cases of the reference Hamiltonian. In the method of {\it self--interference}, we use
\begin{equation}
\hat {\mathcal H} = \frac{\hat p^2}{2m} + \kappa\sin{(k_0\hat x+\pi)}\delta_T(t)
\end{equation}
which differs from the Hamiltonian of the kicked rotor by the phase difference of $\pi$. 

In order to solve Eq. (\ref{eq:theproduct}) for the momentum wave function of the kicked rotor we select the negative diagonal $P=-p$ of the two--dimensional measured probabilities $W_N^{(\theta)}(p;P)$ which yields
\begin{equation}
\phi_N(p)=\frac{1}{\sqrt{2{\mathcal W}_N(0)}}{\mathcal M}_N(p;-p)
.
\end{equation}
Here we have assumed ${\mathcal W}_N(0)\neq0$. Since we have measured the distribution, we already know that value. In case of ${\mathcal W}_N(0)=0$ we use another suitable $P$. Moreover, we have chosen the phase of $\varphi_N(0)$ to vanish. 

In the case of well--separated peaks, that is when the width $\sigma$ of the initial momentum distribution is much smaller than the shift $p_0$, the two momentum distributions $W_N(p)$ and ${\mathcal W}_N(p)$ are identical \cite{bib:remark1}. This property reduces the number of measurements. 

In the {\it holographic method} the reference is provided by the atom in the lower state moving in the absence of any potential, that is 
\begin{equation}
\hat {\mathcal H} = \frac{\hat p^2}{2m}.  
\end{equation}
However, the freely propagated momentum wave function of the lower state is too narrow to cover the state to be reconstructed. For this reason we impart another kick to displace the reference state. 
We can always shift it by integer multiples of $p_0$, that is $P=n p_0$, to cover all significant parts of the momentum wave function. Nevertheless smaller displacements are possible in order to improve the accuracy of the reconstruction scheme. 
 
During the free time evolution the momentum wave function 
$
\varphi_N(p)=\exp(-i\frac{p^2}{2m}\frac{NT}{\hbar})\phi_0(p)
$
only accumulates a phase which yields the measured distribution ${\mathcal W}_N(p)=|\varphi_N(p)|^2/2=|\phi_0(p)|^2/2$. With the help of the inversion formula, Eq. (\ref{eq:theproduct}), we find 
\begin{equation}
\phi_N(p)=\frac{1}{\phi_0^\ast(p+n p_0)}e^{-i N \beta(p+n p_0)}{\mathcal M}_N(p, n p_0).
\label{eq:holo}
\end{equation}

\begin{figure}
\includegraphics[width=8.5cm]{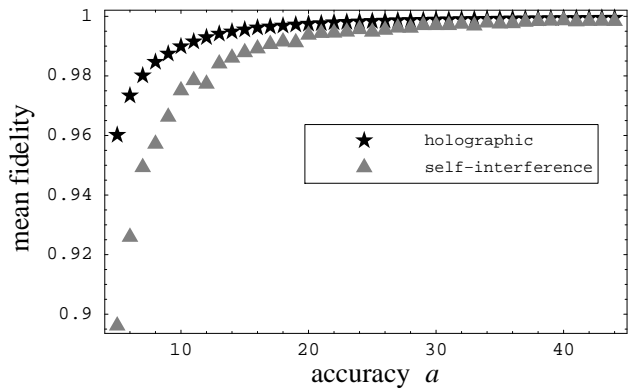}
\caption
{
Comparison between the methods of self--interference and holography based on the fidelity for the state of the kicked rotor after $9$ kicks. 
Shown is the mean fidelity (averaged over 25 realizations) of the reconstructed state versus the accuracy $a$ of the measured distributions. 
The accuracy is defined by $a=1/\Delta W$ with $\Delta W$ being the relative uncertainty of the measured distributions.
}
\label{fig:recfid}
\end{figure}

We now exemplify our reconstruction scheme using a Monte--Carlo
simulation of the holographic method. We propagate the momentum wave function of the kicked rotor with the help of a FFT algorithm. 
Our initial wave function is a Gaussian of width $\sigma$, that is $\braket{p}{\phi_0} \sim \exp(-p^2/4\sigma^2)$.
The parameters of the simulation lead to a comb of 
localized peaks in momentum space. Therefore, we shift the reference state by multiples of $p_0$ in order to reconstruct each single peak and calculate the   distributions $W_N$, $W_N^{(0)}$ and $W_N^{(\frac{\pi}{2})}$. These distributions serve as the weight function for a random number generator which simulates a single measurement event. The distributions emerge from $M$  measurement events, that is measurements of $M$ single atoms. In the final step, these histograms are used to reconstruct an individual peak of the state to be reconstructed with the help of Eq. (\ref{eq:holo}). This procedure is repeated until all significant peaks of the kicked rotor's momentum wave function have been reconstructed. 
In Figs.~\ref{fig:recsim} and \ref{fig:recfid} we show the results of such a simulation for the dimensionless parameters $K \equiv k_0^2 T\kappa/m=14$ and $\kbar\equiv k_0^2 T\hbar/m=15$. The width of the initial momentum wave function is $\sigma=0.1\hbar k_0$.

Figure~\ref{fig:recsim} displays amplitude {\it and} phase of the kicked rotor after $N$ kicks in a comparison between the exact (right column) and the reconstructed (left column) version. We emphasize that the holographic method resolves very well details of the phase portrait.

Figure~\ref{fig:recfid} compares the two methods based on the fidelity defined by the overlap $|\langle \phi_N|\psi_N \rangle|^2$ between the reconstructed state $\ket{\phi_N}$ and the original state $\ket{\psi_N}$ of the kicked rotor after $N$ kicks. The holographic method has a small advantage over the method of self--interference: The fidelity of the reconstructed state is larger, at least for the used parameter regime. 
Moreover, the holographic method relies on fewer measured distributions. Indeed, it needs the distributions $W_N^{(0)}$ and $W_N^{(\frac{\pi}{2})}$ only once per each single peak of the whole state, whereas with the method of self--interference it is necessary to record these distributions for each reconstructed point of the searched wave function. However, the initial state must be known for the holographic method.

We now turn to a brief discussion of a possible experimental implementation of our reconstruction scheme but emphasize that all the ingredients are already in operation. We choose the ($6 {\rm S}_{1/2},\, {\rm F}=3$) and ($6 {\rm S}_{1/2},\, {\rm F}=4$) hyperfine levels of cesium with a level splitting of approximately $9.2{\rm GHz}$ for the reference state $\ket{2}$ and the state $\ket 1$, respectively. Two co-propagating Raman pulses create the initial superposition between them. The frequency $\omega_1$ ($\omega_2$) denotes the transition frequency between $\ket 1$ ($\ket 2$) and an excited electronic state. In the method of self--interference we apply a laser with frequency $(\omega_1+\omega_2)/2$ and $\ket 1$ and $\ket 2$ evolve in potentials which only differ in a $\pi$--phase shift. In the holographic method we pass a laser beam at $\omega_2$ through an electro--optic phase modulator that imposes symmetric sidebands at $\pm\omega_m, \pm 2\omega_m,\dots$ on the carrier. An absorption cell with a Doppler profile smaller than $\omega_m$ strips the carrier but leaves the sidebands unchanged. For example, in cesium we can take $\omega_m/(2\pi)={\rm 4GHz}$. Finally, we split the beam to create the standing wave. Since the sidebands are tuned symmetrically to the red and blue side of the resonance there will be no effect on the reference state but $\ket 1$ will experience a standing wave potential. For example in cesium the dominant term will be detuned $9.2 {\rm GHz}-4  {\rm GHz}=5.2 {\rm GHz}$ from resonance. The same technique can be used for the final kick in the readout stage. However, in this case we create sidebands around $\omega_1$. In this way we can make an accelerating standing wave that will only affect $\ket 2$. Finally, we drive a Raman transition using a $\pi/2$--pulse to detect the atoms in the reference state.

We conclude by summarizing our main results. We have proposed two experimentally feasible methods to measure the wave function of the kicked rotor in amplitude {\it and} phase. Both methods rely on atom interferometry, that is interference between the center--of--mass motions in the two internal states. In this way we bring to light the convoluted behavior of the phases which are at the heart of the phenomena of dynamical localization and the quantum resonances. 

We thank M. Freyberger for many fruitful discussions. The work of MB and WPS is supported by the Deutsche Forschungsgemeinschaft. The work of MGR is supported by the Welch Foundation and the National Science Foundation.

\end{document}